# Dual band wireless transmission over 75-150GHz millimeter wave carriers using frequency-locked laser pairs

**Zichuan Zhou, Amany Kassem, James Seddon, Eric Sillekens, Izzat Darwazeh, Polina Bayvel and Zhixin Liu**

*Department of Electronic and Electrical Engineering, UCL (University College London)*
*Author e-mail address: zczlzz0@ucl.ac.uk*

**Abstract:** We generate and transmit 75-GHz-bandwidth OFDM signals over the air using three mutually frequency-locked lasers, achieving minimal frequency gap between the wireless W and D bands using optical-assisted approaches, resulting in 173.5 Gb/s detected capacity. © 2024 The Author(s)

## 1. Introduction

The surge in new wireless applications, including HD-video streaming, AR/VR, and the Internet of Things, requires faster communication between devices and the Cloud. This has stimulated research in high-capacity mobile front-haul transmission exceeding 100 Gb/s, using analog or digital radio-over-fiber technologies [1]. In deployed systems, however, many base stations do not have access to optical fibers. Instead, high-capacity point-to-point wireless links are needed to connect those base stations to the Cloud to enhance the coverage and capacity of wireless networks.

To achieve this, multiple mm-wave frequency bands, particularly W (75-110 GHz) and D (110-170 GHz) bands are being considered to enhance wireless transmission capacity due to the wider available bandwidth [2,3]. The generation and transmission of mm-wave signals in these high frequency bands have been achieved using both all-electronic and optoelectronic methods, including considerable work exists on developing mm-wave electronic components [4-6], and photodiodes and microwave photonics subsystems optimized for W and D bands [7-8]. While the all-electronic approach exhibits outstanding performance for signals below 70 GHz, it encounters hurdles in mm-wave signal generation, such as reduced power efficiency [6] and enhanced phase noise [9]. Conversely, photonic-assisted approaches, which mix optically modulated signals with a local oscillator (LO) laser on a photodiode, show high efficiency and low noise in mm-wave signal generation. Previous research has demonstrated W and D band signal generation and transmission in separate experiments, including a 125 Gb/s transmission at the W band using a single-in-single-out system [10], and 1.056 Tb/s using 24-GHz-bandwidth with 4 pairs of dual-polarization TRx at D-band, achieving 132 Gb/s per TRx [11].

Nevertheless, the existing optical-assisted mm-wave transmission predominantly use free-running lasers, which precludes simultaneous generation and transmission of both W and D bands signals over the same air interface due to the large drift of the carrier frequencies. Previous work showed locking lasers to a reference etalon or frequency comb for precise carrier frequency generation [12]. Unfortunately, such technique requires extra frequency stabilized comb generator, adding complexity and cost. In this paper, we demonstrate a unique 75-GHz-bandwidth mm-wave signal transmission over 12-cm non-collimated wireless link using the full W and D band spectra, receiving more than 173.5 Gb/s data rate using orthogonal frequency division multiplexing (OFDM) signal. We achieved a minimal frequency gap between W and D bands using mutually frequency locked laser pairs. The locked lasers offer significantly improved phase noise compared to the free-running counterpart, potentially offering enhanced spectrum efficiency using advanced modulation format such as the non-orthogonal spectrally efficient frequency division multiplexing (SEFDM) [13]. The mutual frequency locking eliminates the need for a stable etalon in conventional laser frequency stabilization methods [12], tracing the stability of the mm-wave signals to the same reference oscillator, potentially benefiting sensing and timing functions in future 6G wireless networks.

## 2. Experimental Setup

Fig. 1a shows our experimental setup. The mm-wave signals are generated using three lasers, including an ultra-low linewidth laser (LD1, linewidth of 100 Hz), operating at 1554.82 nm, and two tunable lasers of 5 kHz (LD2) and 80 kHz (LD3) linewidths. The tunable lasers are frequency locked to LD1 with 92.5 and 130 GHz frequency offsets, respectively, as shown in Fig. 1b and 1d. The LD1 is tapped and split using polarization maintaining (PM) splitters before combining with the tapped continuous wave (CW) from LD2 and LD3 for frequency locking. The frequency locking was achieved by detecting the beat tones of the laser pairs and down converted to sub-50 MHz intermediate frequency (IF) before being detected by phase frequency detectors (PFD) [14]. The outputs of the PFDs are fed to two proportional integral (PI) controllers to provide feedback signals that frequency modulate the tunable lasers. We estimate a locking bandwidth close to 100 kHz for both feedback loops, limited by the bandwidth of the piezo-based frequency tuning. Fig. 1c and 1e show the RF spectra of beat tones generated from the LD1-LD2 and LD1-LD3 pairs, respectively, showing a locking bandwidth of about 100 kHz.

The continuous wave (CW) outputs of the LD2 and LD3 are optically amplified and modulated with separate data streams using two 40-GHz LiNbO$_3$ IQ modulators (IQM). Baseband OFDM waveforms of 17.5 GHz and 20 GHz bandwidth are used to drive IQM1 and IQM2 for the respective W-band and D-band signal generation, using synchronized outputs from a 92 GSa/s arbitrary waveform generator (AWG). The OFDM signals contain 256 sub-carriers, each modulated using 16 quadrature amplitude modulation (QAM) format. This yields a subcarrier spacing of 137 MHz and 156 MHz for W-band and D-band signals, respectively. The edge sub-carriers, for each band, are not modulated, resulting in a frequency gap of about 293 MHz between W and D bands. The source data is from a pseudo-random binary sequence (PRBS) of length of $2^{17}$-1. With a clipping ratio of 10 dB, the OFDM digital waveforms obtained have a peak-to-average-power ratio (PAPR) of 12 dB.

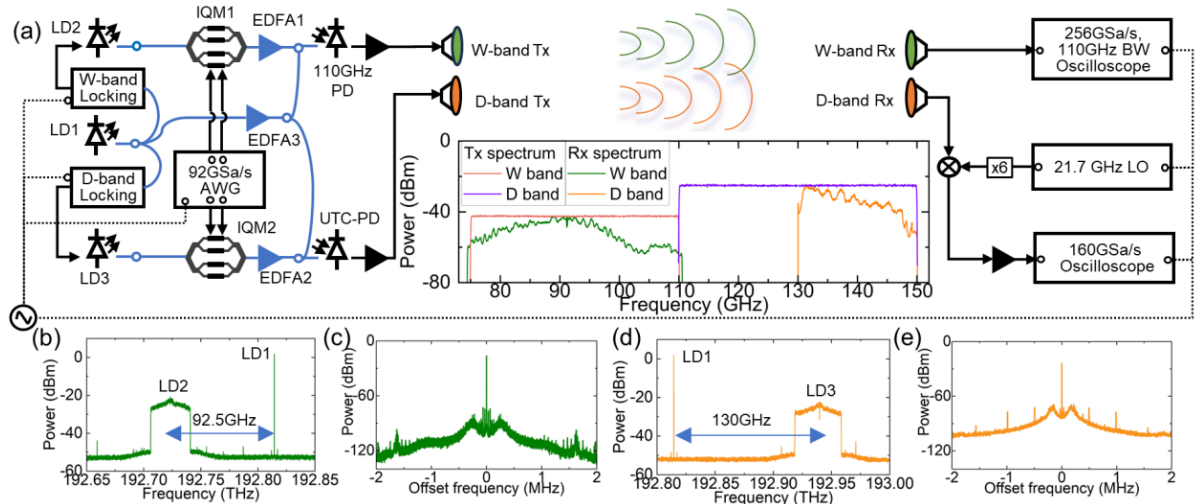

Fig. 1: (a) Experimental setup for simultaneous W and D band wireless transmission, Optical spectrum of (b) modulated LD2 and LD1 and (d) modulated LD3 and LD1, RF spectra of beat tones between (c) LD1 and LD2 and (e) LD1 and LD3

The outputs of the IQMs are subsequently amplified by pre-amplifiers to 9 dBm before combining with the LO CW signal from LD1 using PM couplers. The W-band signal (75-110 GHz) is generated using a 110-GHz PD followed by a 110-GHz RF amplifier, resulting in a 35-GHz-bandwidth OFDM spanning from 75-110 GHz covering entire W-band. The output of the amplifier is connected to a W-band horn antenna with 20-dBi gain (Flann 27240) for the wireless transmission. The D-band signal is generated by detecting the LD1-LD3 pair using a D-band uni-travelling-carrier (UTC) PD followed by a 110-170 GHz RF amplifier, before being transmitted using a 20-dBi horn antenna (Flann 29240). This results in 40-GHz-bandwidth OFDM signal spanning 110-150 GHz.

At the receiver side, the W band signal is detected by a W-band antenna before directly being digitized by a 110-GHz-bandwidtth analog-to-digital converter (ADC) operating at 256 GSa/s (Keysight UXR). The output of D-band antenna is connected to a mixer with an integrated ×6 frequency multiplier at the LO branch. Using a 21.7-GHz seed LO, the D-band signal is down converted to 0.5-17 GHz IF before digitized by a 32-GHz-bandwidth ADC at 160 GSa/s. Although a 40-GHz-bandwidth D-band signal is transmitted, the detection of the D-band signal is limited by the available mm-wave down-converter at the receiver, which only has an IF bandwidth of 17 GHz over 133-150 GHz. The inset in Fig. 1a shows the measured power spectrum of both transmitted and received W band and D band signal at 100 MHz resolution. The pink and purple lines show the digital spectra of the transmitted signals, centered at 92.5 GHz and 130 GHz, respectively. The green and orange lines show the measured spectra from the receivers. The W band signal exhibits a 10 dB roll-off in 100-110 GHz due to the limited bandwidth of photodetector and RF amplifier. The received D band signal power significantly reduces beyond 147 GHz due to limited IF bandwidth of the mm-wave down-convertor. The digitized signals were offline-processed using digital signal processing (DSP) methods reported in [15], and the bit loading [16] of each band is estimated based on the error vector magnitude of the signals assuming 15.5% overhead soft-decision forward error correction (SD-FEC) of BER of 2.2e-2 [17].

## 3. Experimental Results

Fig. 2a shows the measured phase noise of the carrier signals generated by beating CW of the locked LD1-LD2 (green solid curve) and LD1-LD3 (orange dashed curve), respectively. The black dotted curve shows the LD1-LD2 beat when the lasers are free running. The locked lasers show significant improvement of phase noise within the locking bandwidth. The locked LD1-LD3 pair exhibits worse phase noise due to higher fundamental linewidth of the LD3.

Fig. 2b shows the measured SNR of demodulated OFDM sub-carriers for the W and D band signals, shown in green diamond and orange star markers. Both bands exhibit similar average SNR close to 12dB, limited by wireless link loss. Note that the SNRs of sub-carriers located at edges of W and D band are lower compared to center sub-carriers, which matches with the power spectrum shown by inset in Fig. 1a. Fig. 2c and 2d show the 16QAM constellation diagram of selected sub-carriers (shown by arrow in Fig. 2b). The SNRs (or EVMs) of the subcarriers determine the achievable order of modulation, as shown in Fig. 2e. The best-performing W-band subcarriers can be modulated up to 32QAM (5 bits/symbol). Due to reduced power, caused by sharp frequency roll-off of components, sub-carriers from 100-110GHz can only be modulated using QPSK (2 bits/symbol) or BPSK. On the other hand, the D-band subcarriers mostly support 16 QAM (4 bits/symbol) due to the relatively smooth roll-off of mm-wave down-convertor.

The estimated capacity using EVM-based bit-loading is 106 and 67.5 Gb/s, for the detected W and D band signals, respectively. This results in an aggregated capacity of 173.5 Gb/s, corresponding to a net rate of just over 150 Gb/s. Finally, we note that the transmission distance is limited by the unfocused beams, which can be extended to more than 100 meters with lenses [11].

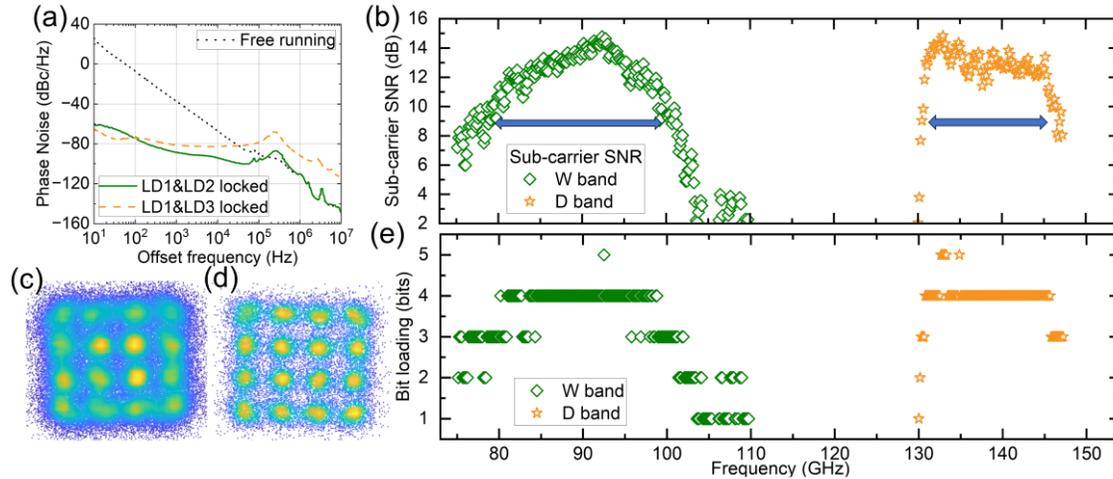

Fig. 2: (a) Phase noise measurement of RF carrier, (b) SNR of demodulated OFDM sub-carriers, 16QAM constellation diagram of (c) W band, (d) D band, (e) bit loading of sub-carriers.

## 4. Conclusion

We demonstrate simultaneous W and D band wireless transmission using frequency-locked optical heterodyne methods, achieving a total of 75 GHz wireless transmission bandwidth and a detected capacity of 173.5 Gb/s. Our technology will benefit the high-capacity transmission requirements between base stations and can potentially benefit high-capacity needs in in-door wireless communications and intra-data center interconnections.

The authors acknowledge EPSRC grants EP/V051377/1, EP/W026252/1 and EP/R035342/1 for funding support.